\documentclass[aps,prb,superscriptaddress,citeautoscript,twocolumn,longbibliography]{revtex4-1}
\usepackage{graphicx}
\usepackage{amsfonts}
\usepackage{amsmath}
\usepackage{amssymb}
\usepackage{bm}

\usepackage{bbold}
\usepackage{mathtools}
\usepackage{dcolumn}
\usepackage{hyperref}
\usepackage[usenames]{color}
\usepackage{subcaption}
\usepackage{xfrac}

\usepackage{physics}
\usepackage{units}

\usepackage{tablefootnote}

\usepackage{xcolor}

\AtBeginDocument{%
	\newwrite\bibnotes
	\def\bibnotesext{Notes.bib}
	\immediate\openout\bibnotes=\jobname\bibnotesext
	\immediate\write\bibnotes{@CONTROL{REVTEX41Control}}
	\immediate\write\bibnotes{@CONTROL{%
			apsrev41Control,author="08",editor="1",pages="1",title="0",year="1"}}
	\if@filesw
	\immediate\write\@auxout{\string\citation{apsrev41Control}}%
	\fi
}%

\usepackage{ragged2e}
\DeclareCaptionJustification{justified}{\justifying}
\def\beq{\begin{equation}}
\def\eeq{\end{equation}}

\newcommand{\out}[1]{{}}

\newcommand{\fref}[1]{Fig.~\ref{#1}}

\newcommand{\kvec}[1]{{\mathbf{#1}}}

\begin{document}

\title{Probing sliding ferroelectricity in bilayer T$_\mathrm{d}$-WTe$_2$ with high-harmonic generation}

\author{Elias Greil}
\address{Institute of Theoretical and Computational Physics, TU Graz, NAWI Graz, Petersgasse 16, 8010 Graz, Austria}

\author{Alba de las Heras}
\address{Max Planck Institute for the Structure and Dynamics of Matter, Center for Free Electron Laser Science, Luruper Chaussee 149, 22761 Hamburg, Germany}

\author{Angel Rubio}
\address{Max Planck Institute for the Structure and Dynamics of Matter, Center for Free Electron Laser Science, Luruper Chaussee 149, 22761 Hamburg, Germany}
\address{Center for Computational Quantum Physics, Flatiron Institute, New York, NY 10010, USA}
\address{Nano-Bio Spectroscopy Group, and ETSF, Universidad del País Vasco UPV/EHU- 20018 San Sebastián, Spain}

\author{Anna Galler}
\email{anna.galler@tugraz.at}
\address{Institute of Theoretical and Computational Physics, TU Graz, NAWI Graz, Petersgasse 16, 8010 Graz, Austria}
\address{Max Planck Institute for the Structure and Dynamics of Matter, Center for Free Electron Laser Science, Luruper Chaussee 149, 22761 Hamburg, Germany}

\begin{abstract}
High-harmonic generation is a sensitive all-optical probe of symmetry and electron dynamics in solids. Here, we use first-principles time-dependent density functional theory (TDDFT) to study high-harmonic generation in T$_\mathrm{d}$-WTe$_2$, a two-dimensional semimetal with switchable out-of-plane ferroelectric polarization driven by interlayer sliding. We show that the mirror-symmetry breaking underlying the ferroelectric state produces robust signatures in polarization-resolved high-harmonic spectra, enabling optical identification of the polarization state. By incorporating interlayer shear motion in coupled electron-lattice TDDFT simulations, we further show that the 0.24 THz shear mode is slow enough to remain effectively decoupled from the ultrafast electronic response responsible for harmonic emission. Our results establish high-harmonic spectroscopy as a non-invasive probe of sliding ferroelectricity and lattice symmetry in two-dimensional quantum materials.\end{abstract}

\maketitle

\section{Introduction}
Sliding ferroelectrics are bilayer or multilayer van der Waals (vdW) materials that exhibit robust out-of-plane ferroelectric (FE) polarization. This polarization arises from specific stacking configurations that break spatial inversion symmetry and can be reversibly switched via interlayer sliding. Similar to conventional bulk ferroelectrics, these systems are promising for applications such as non-volatile memory and optoelectronic devices. Moreover, in contrast to their bulk counterparts, two-dimensional (2D) ferroelectrics offer several advantages, including reduced size, mechanical flexibility, and lower switching fields.\cite{Zhang2025review}

Starting from the initial proposal in 2017\cite{Li2017}, a growing number of sliding ferroelectrics have been identified, including bilayer graphene\cite{zheng2020graphene}, boron nitride\cite{yasuda2021bn,wang2024bevel,Wong2025}, InSe$_2$\cite{hu2019,sui2023}, and transition-metal dichalcogenides (TMDs) such as MoS$_2$, MoSe$_2$, WS$_2$ and WSe$_2$\cite{rogee2022,wang2022,deb2022cumulative}. In particular, bilayer T$_\mathrm{d}$-MoTe$_2$ and T$_\mathrm{d}$-WTe$_2$ have attracted attention due to the interplay of ferroelectricity and nontrivial band topology\cite{fei2018switch,sharma2019semimetal,xiao2020berry,de2021direct}. Their non-centrosymmetric, orthorhombic T$_\mathrm{d}$ phase (space group $Pnm2_1$) hosts type-II Weyl semimetallic behaviour in the bulk\cite{kim2017struct,sie2019ultrafastdyn,guan2021weyldyn,hein2020sheardyn,jindal2023ferrosuper}. T$_\mathrm{d}$-WTe$_2$ remains semimetallic down to the bilayer limit, and transitions to a quantum spin Hall insulator in the monolayer\cite{tang2017monotopo,fei2017monotopo,xu2018monophotoberry,sajadi2018monosuper}. Although bilayer T$_\mathrm{d}$-WTe$_2$ is metallic in-plane, its atomic thickness allows the out-of-plane FE polarization to persist despite electronic screening. The resulting polarity is robust, vanishing only above \unit[350]{K}, and can be switched by an electric field applied along the out-of-plane direction\cite{fei2018switch}. Notably, the switching mechanism does not involve vertical ionic displacement, but instead arises from interlayer sliding\cite{yang2018dft}, consistent with other sliding ferroelectrics such as boron nitride and MoS$_2$.\cite{Li2017}

This sensitivity of layered materials to shear and twist degrees of freedom has also been demonstrated dynamically. In WTe$_2$, ultrafast electron diffraction showed that terahertz pulses can drive large-amplitude interlayer shear strain, inducing a transient symmetry change toward a centrosymmetric, topologically trivial phase and enabling ultrafast creation or annihilation of Weyl points.\cite{sie2019ultrafastdyn} More recently, combined ultrafast electron diffraction with atomic-scale simulations revealed coherent photoinduced twisting and untwisting of moiré superlattices.\cite{duncan2025photoinduced} Together, these studies establish interlayer shear, sliding, and twist as active structural coordinates for controlling ferroelectric, moiré, and topological properties on ultrafast timescales. They further point to a broader opportunity: using external stimuli--such as electric fields, terahertz pulses, and optical excitation--to manipulate vdW stacking and dynamically tune symmetry, polarization, and topology in atomically thin materials.

Building on these advances, time-resolved pump–probe experiments have directly demonstrated ultrafast symmetry switching in T$_\mathrm{d}$-WTe$_2$\cite{sie2019ultrafastdyn} and T$_\mathrm{d}$-MoTe$_2$\cite{Zhang2019switch}. In these experiments, terahertz or optical pulses are used to coherently excite a shear phonon mode, while the resulting lattice symmetry is monitored via time-resolved second harmonic generation (SHG) and ultrafast electron diffraction. Both studies show that sufficiently intense pump pulses drive the materials, through the excitation of a shear phonon, from the non-centrosymmetric T$_\mathrm{d}$ phase to the centrosymmetric 1T$^\prime$ phase.

While SHG is a well-established technique for distinguishing lattice structures with and without inversion symmetry, strong-field phenomena such as ultrafast photocurrents\cite{higuchi2017light,xu2018monophotoberry,mciver2012photo,Galler2025} and high-harmonic generation (HHG)\cite{goulielmakis2022high} are also emerging as powerful spectroscopic probes. In particular, HHG has been shown to encode signatures of ultrafast electron dynamics\cite{baudisch2018ultrafast,uzan2020attosecond}, electronic band structure\cite{vampa2015,Lanin2017,Tancogne2017,kim2025hhg}, many-body interactions\cite{silva2018high,murakami2018mott}, and the interplay between Berry curvature and lattice symmetries in topological materials\cite{luu2018berry,neufeld2023topo,zhang2024jonesmatrix,delasHeras2026}. Moreover, 2D materials provide a particularly favourable platform for HHG, as reduced absorption and negligible phase-matching constraints enable more efficient harmonic generation compared to bulk crystals.\cite{Liu2017,Yoshikawa2017,Tancogne2018, Yoshikawa2019,Tyulnev2024,kim2025hhg} 

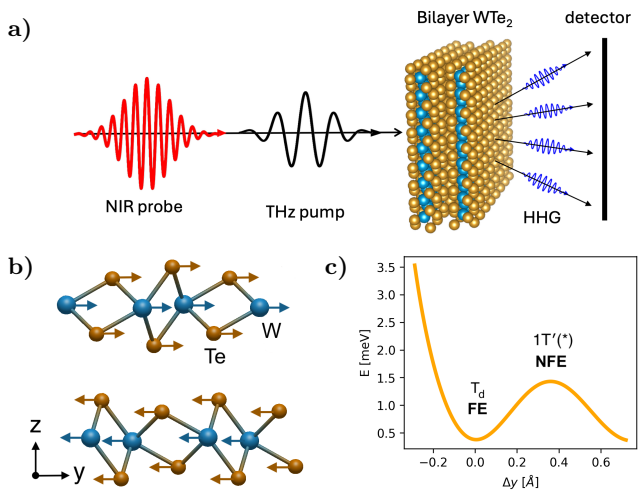
\begin{figure}
    
    \includegraphics[width=1\linewidth]{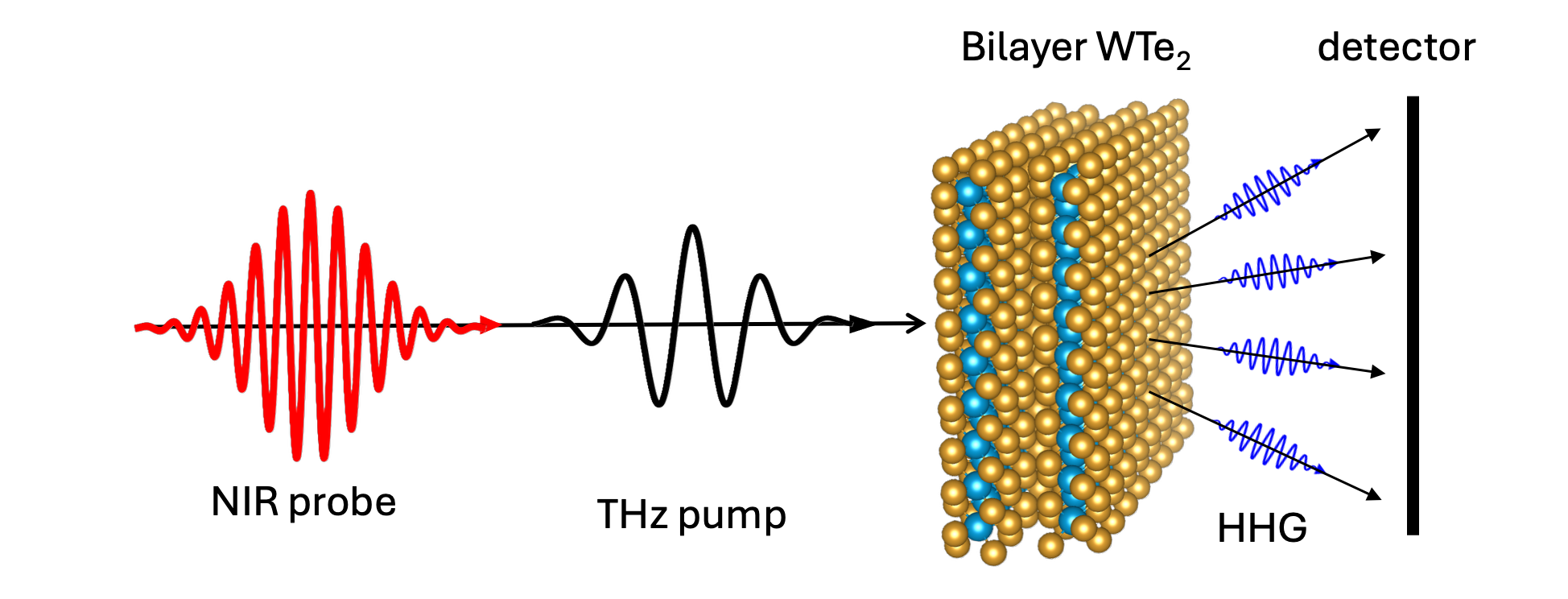}
    
    \begin{minipage}{0.45\linewidth}
    \includegraphics[width=1\linewidth]{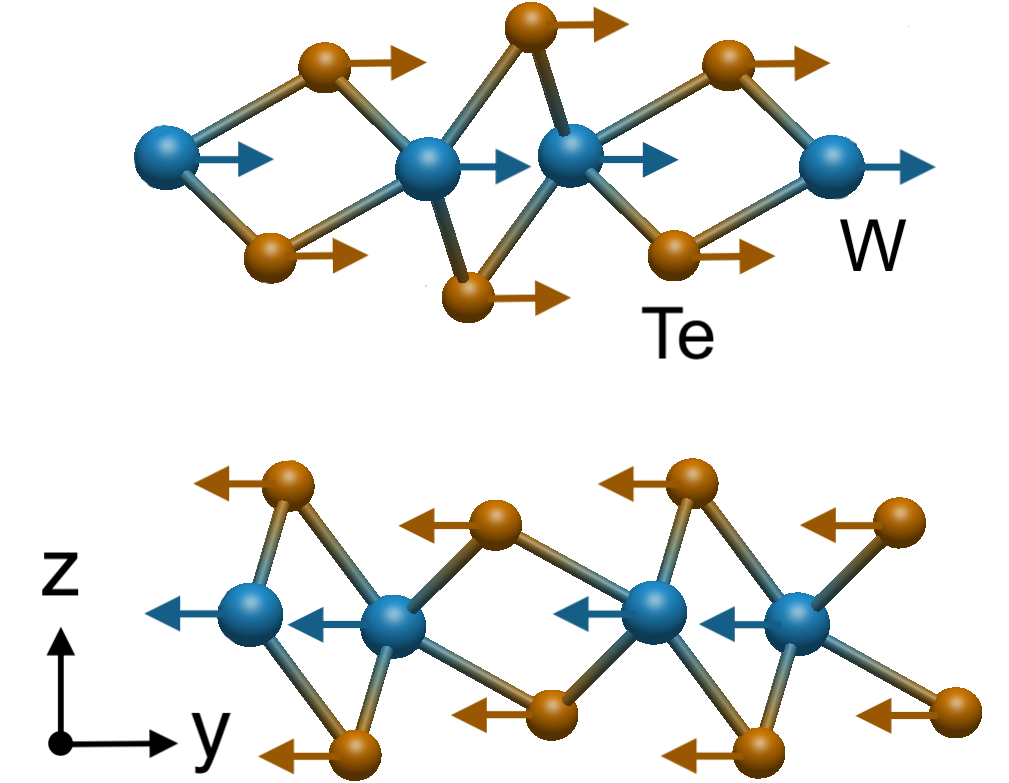}

    \end{minipage}
    \hfill
    \begin{minipage}{0.5\linewidth}
    \includegraphics[width=0.9\linewidth]{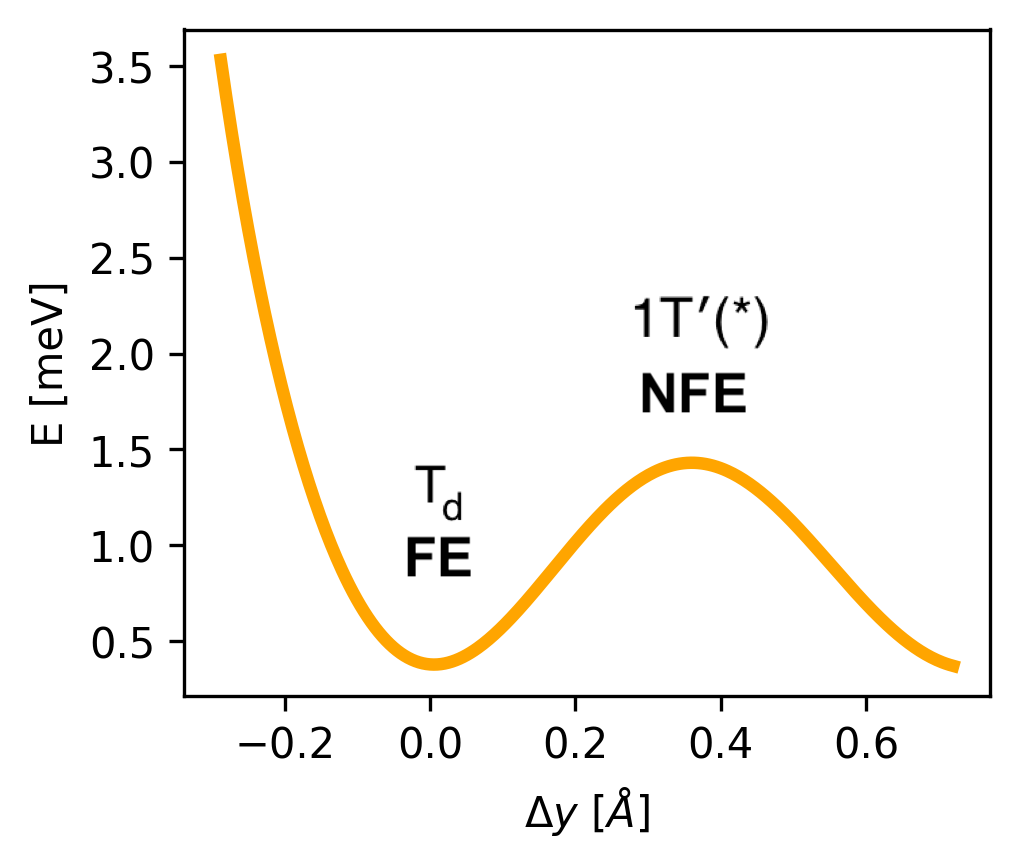}
    \end{minipage}
    \begin{picture}(0,0)
    \put(-250,130){\textbf{a)}}
    \put(-250,40){\textbf{b)}}
    \put(-130,40){\textbf{c)}}
    \end{picture}
        \caption{\textbf{Schematic of the THz pump - NIR probe setup for ferroelectric switching.} (a) A THz pump pulse excites coherent phonons in bilayer T$_\mathrm{d}$-WTe$_2$, followed by a time-delayed NIR pulse, which drives HHG to probe the lattice and electronic structure.  (b) Dynamics of the \unit[0.24]{THz} interlayer shear phonon. (c) Potential energy as a function of interlayer shear displacement, calculated using the nudged elastic band (NEB) method within DFT (see Methods). The THz pulse can drive a transition from the FE T$_\mathrm{d}$ phase to a non-ferroelectric (NFE) intermediate state.\cite{sie2019ultrafastdyn} }
    \label{fig:experiment}
\end{figure}

In this work, we investigate whether HHG encodes information about the FE polarization in a 2D sliding ferroelectric. More broadly, we assess the potential of HHG as an all-optical spectroscopic tool capable of directly probing lattice dynamics and FE polarization on ultrafast timescales. We theoretically consider a pump–probe scheme, depicted in \fref{fig:experiment}a). In this setup, bilayer T$_\mathrm{d}$-WTe$_2$ is excited by a THz pump pulse. As in many quantum materials\cite{liu2022coherent,jiang2023coherent, notter2025dynamical}, such intense ultrafast pump pulses can drive coherent phonon dynamics that transiently modify the lattice and electronic structure. In T$_\mathrm{d}$-WTe$_2$, an interlayer shear phonon mode at \unit[0.24]{THz} is launched, corresponding to a dynamic interlayer displacement along the $y$ direction (\fref{fig:experiment}b). Strong THz excitation can induce large-amplitude shear motion, driving a structural transition from the FE T$_\mathrm{d}$ phase to a non-ferroelectric (NFE) intermediate state (\fref{fig:experiment}c), which we denote as 1T$^\prime(^*)$, following the notation in Ref.~\onlinecite{sie2019ultrafastdyn}.  Since the FE switching is intrinsically connected to the evolution of the lattice and electronic structure, we aim to identify its characteristic fingerprints in the HHG spectra driven by a time-delayed near-infrared (NIR) probe pulse.

\section{Results}
 
\subsection*{Signatures of ferroelectricity in HHG}
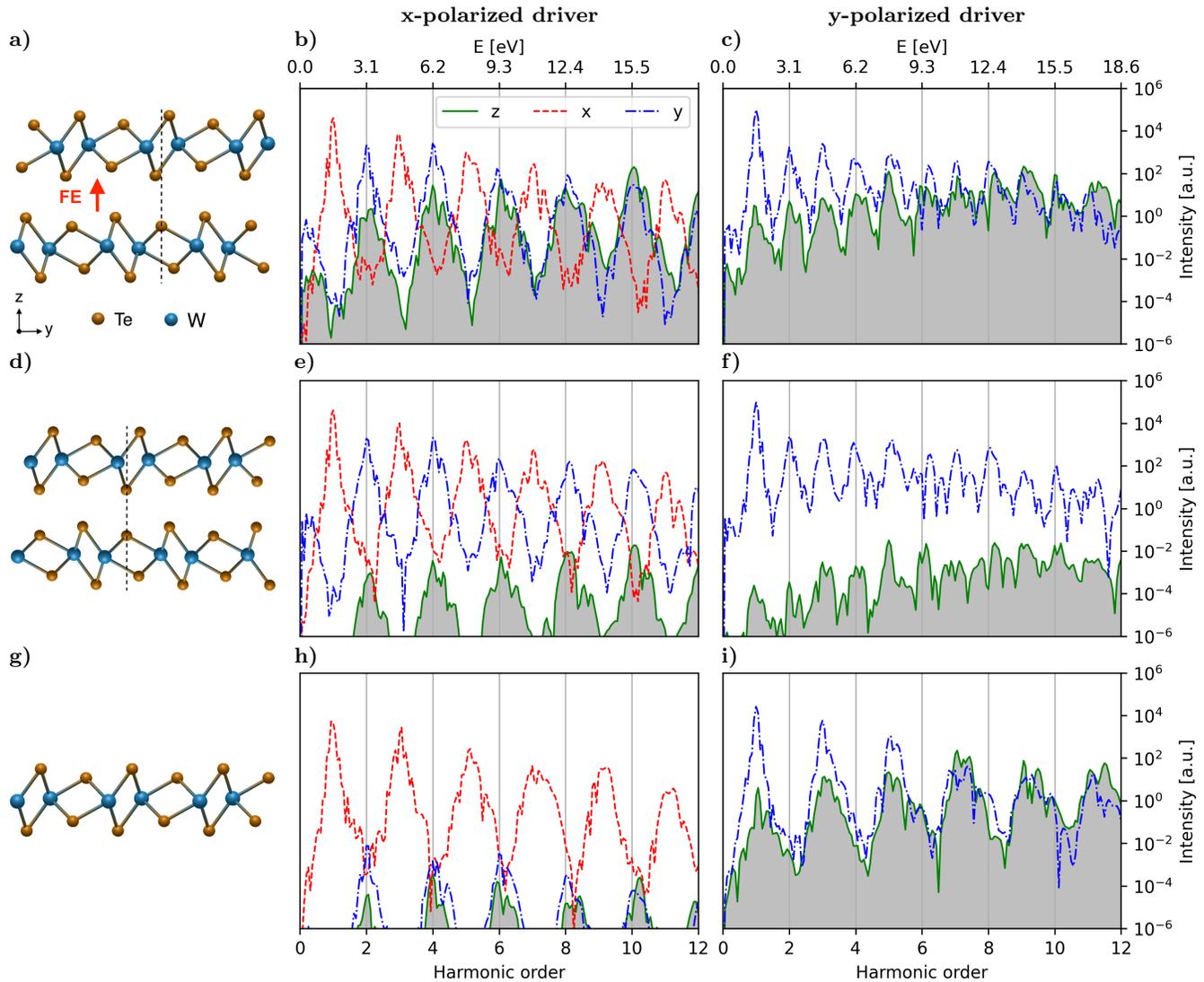
\begin{figure*}
    \centering
    \includegraphics[width=1\linewidth]{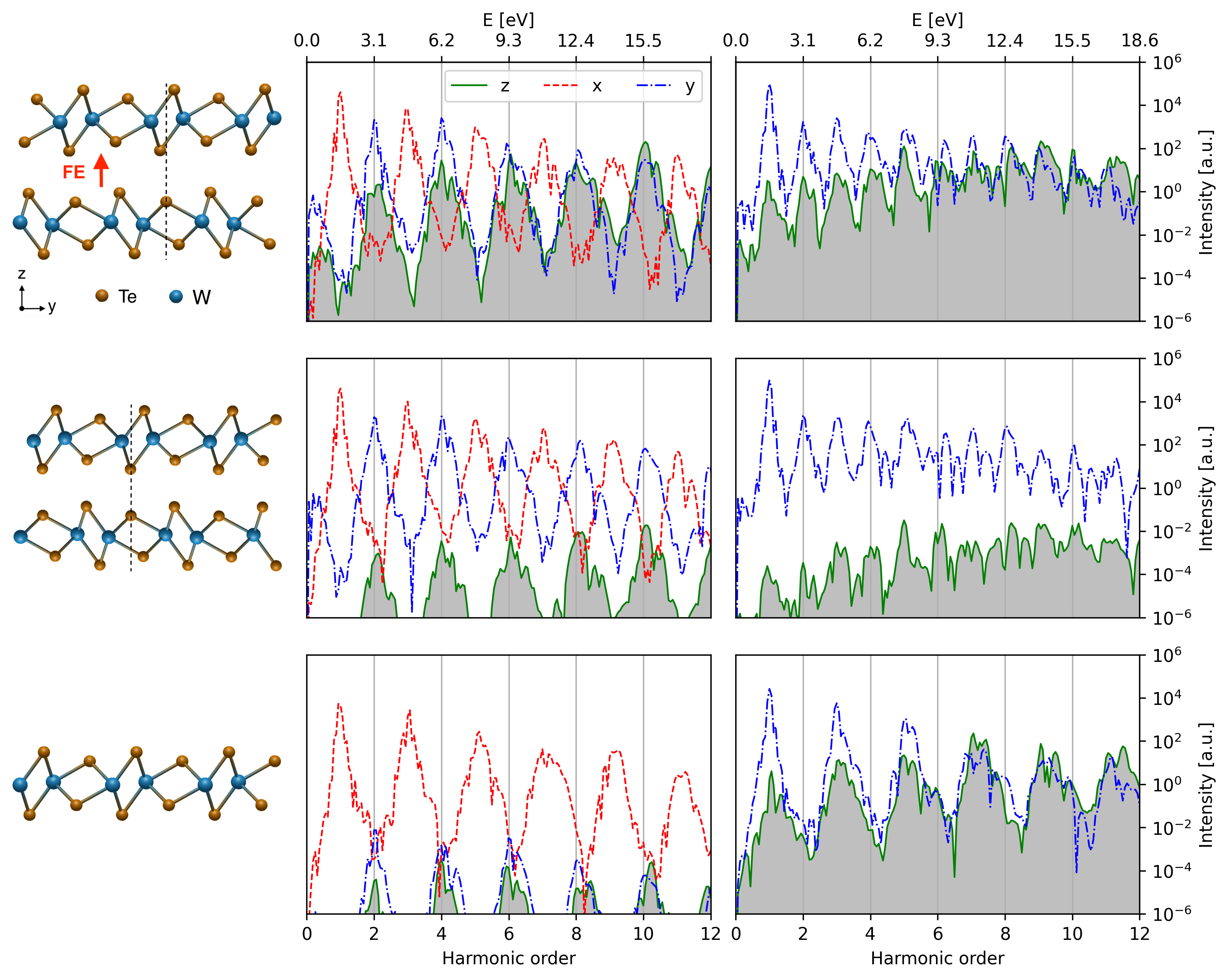}
    \begin{picture}(0,0)
    \put(-250,410){\textbf{a)}}
    \put(-130,410){\textbf{b)}}
    \put(50,410){\textbf{c)}}
    \put(-250,274){\textbf{d)}}
    \put(-130,274){\textbf{e)}}
    \put(50,274){\textbf{f)}}
    \put(-250,150){\textbf{g)}}
    \put(-130,150){\textbf{h)}}
    \put(50,150){\textbf{i)}}
    \put(-85,420){\textbf{x-polarized driver}}
    \put(95,420){\textbf{y-polarized driver}}
    \end{picture}
    \caption{\textbf{Signatures of sliding ferroelectricity in HHG.} (a) The polar T$_\mathrm{d}$ structure of bilayer WTe$_2$ exhibits FE polarization in the out-of-plane ($z$) direction. (b) HHG spectrum obtained for a 16-cycle driving laser pulse with a wavelength of \unit[800]{nm}, linearly polarized along the $x$ direction, and an intensity of \unit[1]{TW/cm$^2$}. (c) Corresponding HHG spectrum for a driving laser polarized along the $y$ direction. (d–f) Same as in (a–c), but for the non-ferroelectric (NFE) intermediate structure, obtained from T$_\mathrm{d}$ via an interlayer shear displacement along $y$. The $z$-polarized harmonics (shaded in grey) are strongly suppressed. (g–i) For a full comparison, we also show the HHG spectra from monolayer WTe$_2$. }
    \label{fig:HHG_XY}
\end{figure*}

We begin by decomposing the proposed pump–probe experiment into a series of static snapshots. In a first step, we analyze HHG in bilayer WTe$_2$ for static ionic configurations. The relevant structural reference points are the FE T$_\mathrm{d}$ phase and the NFE intermediate 1T$^\prime(^*)$ state, shown in \fref{fig:HHG_XY}a) and \fref{fig:HHG_XY}d), respectively. Notably, the 1T$^\prime(^*)$ structure can be obtained from  T$_\mathrm{d}$ via an interlayer shear displacement along the $y$ direction, defining bilayer WTe$_2$ as a sliding ferroelectric.

Bilayer WTe$_2$ is modeled using density functional theory (DFT) within PBE\cite{PhysRevLettPBE}, with vdW interactions included via the DFT-D3 method with Becke–Johnson damping\cite{dft-d3}. The T$_\mathrm{d}$ structure is fully relaxed, and the nudged elastic band (NEB) method\cite{henkelman2000} is used to determine the minimum-energy pathway between the FE T$_\mathrm{d}$ state, the intermediate NFE 1T$^\prime(^*)$ state, and the symmetry-equivalent state with reversed polarization, generated by a mirror operation across the interlayer $xy$ plane. Computational details are provided in the Methods section. 
The resulting energy profile along the switching pathway is shown in \fref{fig:experiment}c. A small interlayer shear displacement of \unit[0.72]{\AA} induces switching of the FE polarization. The associated energy barrier is approximately \unit[1]{meV}, indicating that this sliding-driven mechanism constitutes a realistic switching pathway.

We next compute the HHG response along the FE switching pathway. To this end, the structures obtained from the NEB calculations are exposed to a 16-cycle NIR pulse with a wavelength of \unit[800]{nm} and a peak intensity of \unit[1]{TW/cm$^2$}. The time-dependent electron dynamics is simulated using time-dependent density functional theory (TDDFT), as implemented in the Octopus code\cite{Octopus2020}. Further details of the TDDFT simulations are provided in the Methods section. 

In \fref{fig:HHG_XY}, we show the computed HHG spectra for the main structures: the FE T$_\mathrm{d}$ phase (first row) and the intermediate NFE phase (second row). For comparison, results for a T$_\mathrm{d}$ monolayer are included (third row). We display the polarization-resolved HHG spectra considering a linearly polarized NIR driving field oriented along the in-plane $x$ (second column) and $y$ (third column) directions. 

For a driving field polarized along $x$ (\fref{fig:HHG_XY}b), the HHG spectrum from the FE phase presents $x$, $y$, and $z$ polarization components. Harmonics polarized parallel to the driving field ($x$) are purely odd, whereas those polarized perpendicular to it ($y$ and $z$) are even.
In contrast, for a driving field polarized along $y$ (\fref{fig:HHG_XY}c), both even and odd harmonics are emitted in the $y$ and $z$ directions, while no harmonic emission is observed along $x$.

A comparison with the HHG spectra of the NFE intermediate state (\fref{fig:HHG_XY}d–f) reveals a pronounced suppression of harmonics polarized along the $z$ direction. This contrast is especially important because $z$ corresponds to the out-of-plane direction in which the FE polarization develops in bilayer T$_\mathrm{d}$-WTe$_2$. Therefore, when bilayer WTe$_2$ is driven within the $xy$ plane, the emergence $z$-polarized harmonics provides a clear spectroscopic fingerprint of the FE polarization.

For completeness, we also consider the monolayer case (\fref{fig:HHG_XY}g–i), where only odd-order harmonics are observed. 
The harmonic response of the FE, NFE, and monolayer structures under different driving polarizations--linear polarization along the $x$, $y$, and $z$ directions--are summarized in Table \ref{tab:allowed_harmonics}.

\begin{table}
    \caption{Summary of the HHG response for the ferroelectric T$_\mathrm{d}$, non-ferroelectric 1T$^\prime(^*)$, and monolayer structures. The polarization of the fundamental corresponds to that  of the driving NIR field.}
    \label{tab:allowed_harmonics}
    \begin{tabular}{l|c|c|c|c|c|c}
        \hline
        Fundamental &\multicolumn{2}{c}{$x$} & \multicolumn{2}{c}{$y$} & \multicolumn{2}{c}{$z$}\\
        \hline
        Harmonics& \; odd \; & \; even \; & \; odd \; & \; even \; & \; odd \; & \; even \; \\
        \hline
        T$_\mathrm{d}$  & $x$ & $y$,$z$ & $y$,$z$ & $y$,$z$ & $y$,$z$ & $y$,$z$ \\
        \hline
        1T$^\prime(^*)$ & $x$ & $y$ & $y$ & $y$ & $z$ & $y$\\
        \hline
        monolayer & $x$ & & $y$,$z$ & & $y$,$z$&\\
        \hline
    \end{tabular}
\end{table}

Having established that the out-of-plane $z$-polarized harmonics vanish--or are at least strongly suppressed--in the NFE phase, we now examine how their intensity evolves along the FE switching pathway obtained from the nudged elastic band (NEB) calculations. This pathway connects the FE T$_\mathrm{d}$ structure of bilayer WTe$_2$ to its mirrored configuration with opposite polarization, passing through the intermediate NFE 1T$^\prime(^*)$ state.

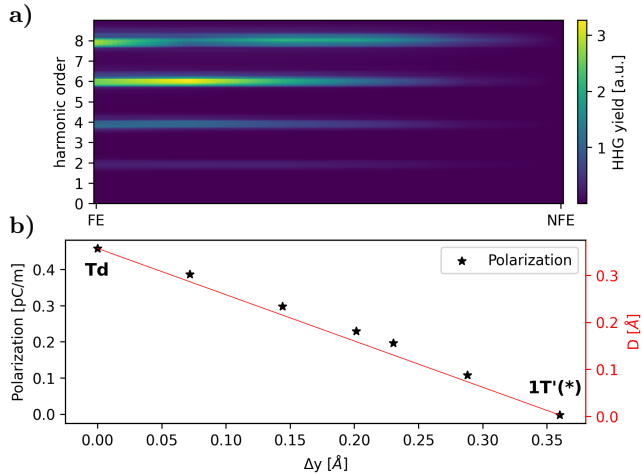
\begin{figure}
    \centering
    \includegraphics[width=1\linewidth]{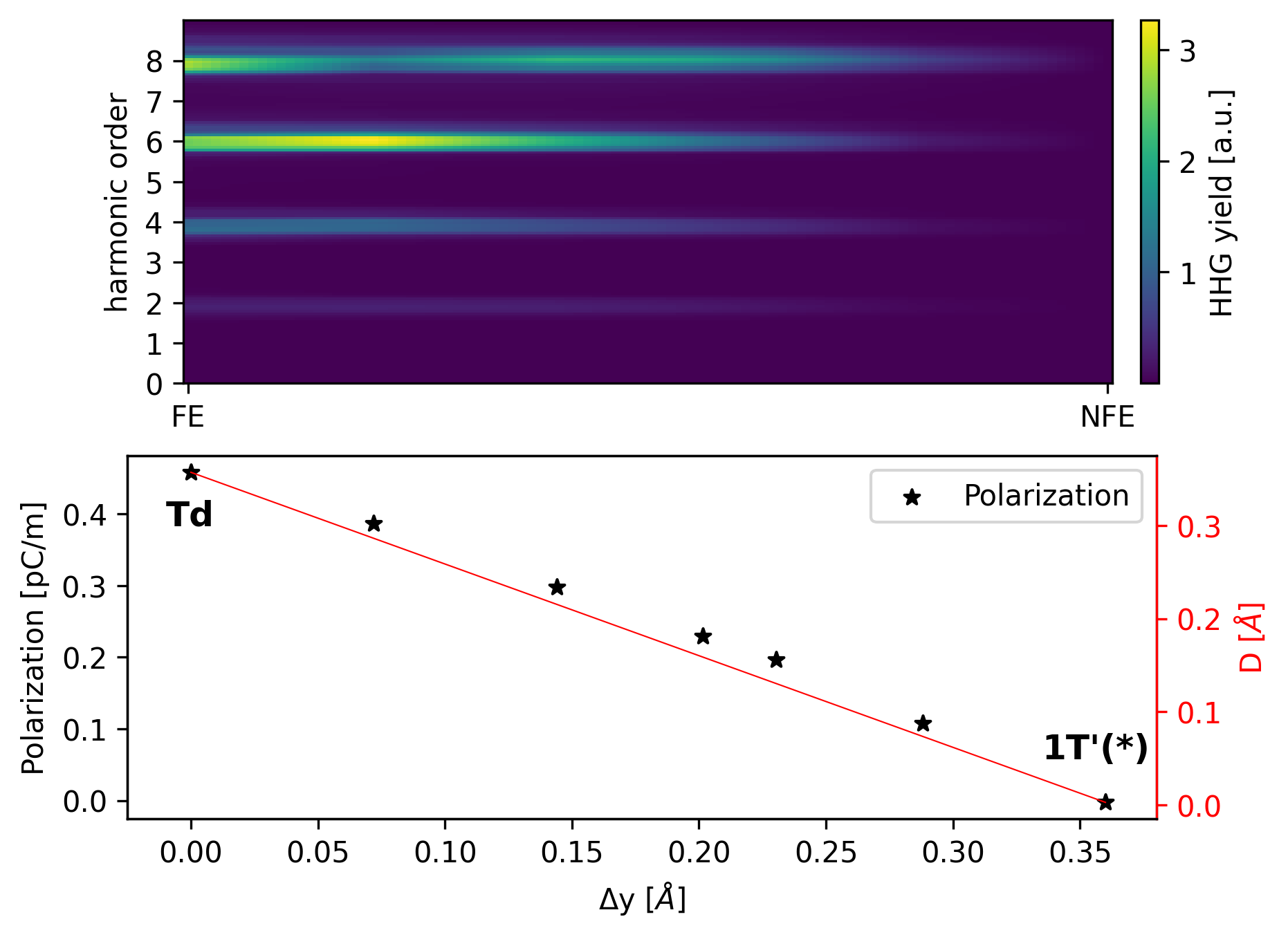}
    \begin{picture}(0,0)
    \put(-120,185){\textbf{a)}}
    \put(-120,105){\textbf{b)}}
    \end{picture}
    \caption{\textbf{HHG response along the FE switching pathway.} 
    (a) Harmonic yield of the $z$-polarized harmonics as a function of the shear displacement $\Delta y$. The HHG yield is computed for an $x$-polarized driving field with a wavelength of \unit[800]{nm} and an intensity of \unit[1]{TW/cm$^2$}. The $z$ harmonics progressively vanish as the shear displacement approaches the NFE 1T$^\prime(^*)$ structure. (b) The ferroelectric polarization is directly proportional to the deviation from the $xy$ glide-plane symmetry, which is present only in the 1T$^\prime(^*)$ structure.  The deviation from this symmetry is quantified by a mean absolute displacement measure $D$ (see Methods).   }
    \label{fig:SYM_HHG}
\end{figure}

In \fref{fig:SYM_HHG}b, we show how the FE polarization progressively decreases along the shear displacement in the $y$ direction, vanishing at the NFE 1T$^\prime(^*)$ structure. The corresponding evolution of the $z$-polarized component in HHG is presented in \fref{fig:SYM_HHG}a, along the same pathway. Considering an $x$-polarized driving field, only even-order harmonics are polarized along $z$. Notably, their intensity decreases continuously and ultimately disappears as the system approaches the NFE state.
A key question, which we will address in the next section, concerns the origin of these 
$z$-polarized harmonics and their suppression in the NFE phase.  
 
\subsection*{A matter of symmetries}
HHG is extremely sensitive to the symmetries of both the driving laser field and the underlying material system\cite{Alon1998,Neufeld2019}. In isotropic, highly symmetric targets, such as atoms, a linearly polarized driving field is known to generate only odd-order harmonics, with the emitted radiation preserving the polarization of the driver. Within the dipole approximation, even-order harmonics are strictly forbidden in centrosymmetric systems and can appear only when  inversion symmetry  is broken. 

In crystalline solids, however, the symmetry constraints are richer. Beyond inversion symmetry, rotational symmetries, mirror planes, and glide or screw rotations can impose additional selection rules that determine which harmonic orders are allowed or suppressed in each polarization channel\cite{Liu2017,Neufeld2019,delasHeras2026}.Consequently, the polarization of the emitted harmonics need not coincide with that of the driving field. Instead, the emitted HHG signal encodes the symmetry of the lattice and electronic stucture, as clearly observed in \fref{fig:HHG_XY} for WTe$_2$.
A systematic framework connecting harmonic orders, emission polarization, and crystal symmetries has recently been developed in Ref.~\onlinecite{zhang2024jonesmatrix}, providing a useful basis for interpreting the polarization-resolved HHG spectra discussed here. This Jones matrix formalism for HHG in crystalline solids reveals how the crystal symmetry controls the allowed harmonic orders and their polarization. In the Methods section, we derive the Jones matrices for all relevant crystal structures and driving-field polarizations considered in this work, and demonstrate that the features observed in Fig.~\ref{fig:HHG_XY} can be consistently interpreted within this framework. In the following, we highlight the most important findings.

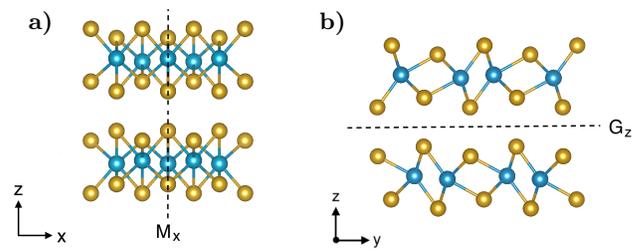
\begin{figure}
    \centering
    
    \begin{minipage}{0.49\linewidth}
        \includegraphics[width=1\linewidth]{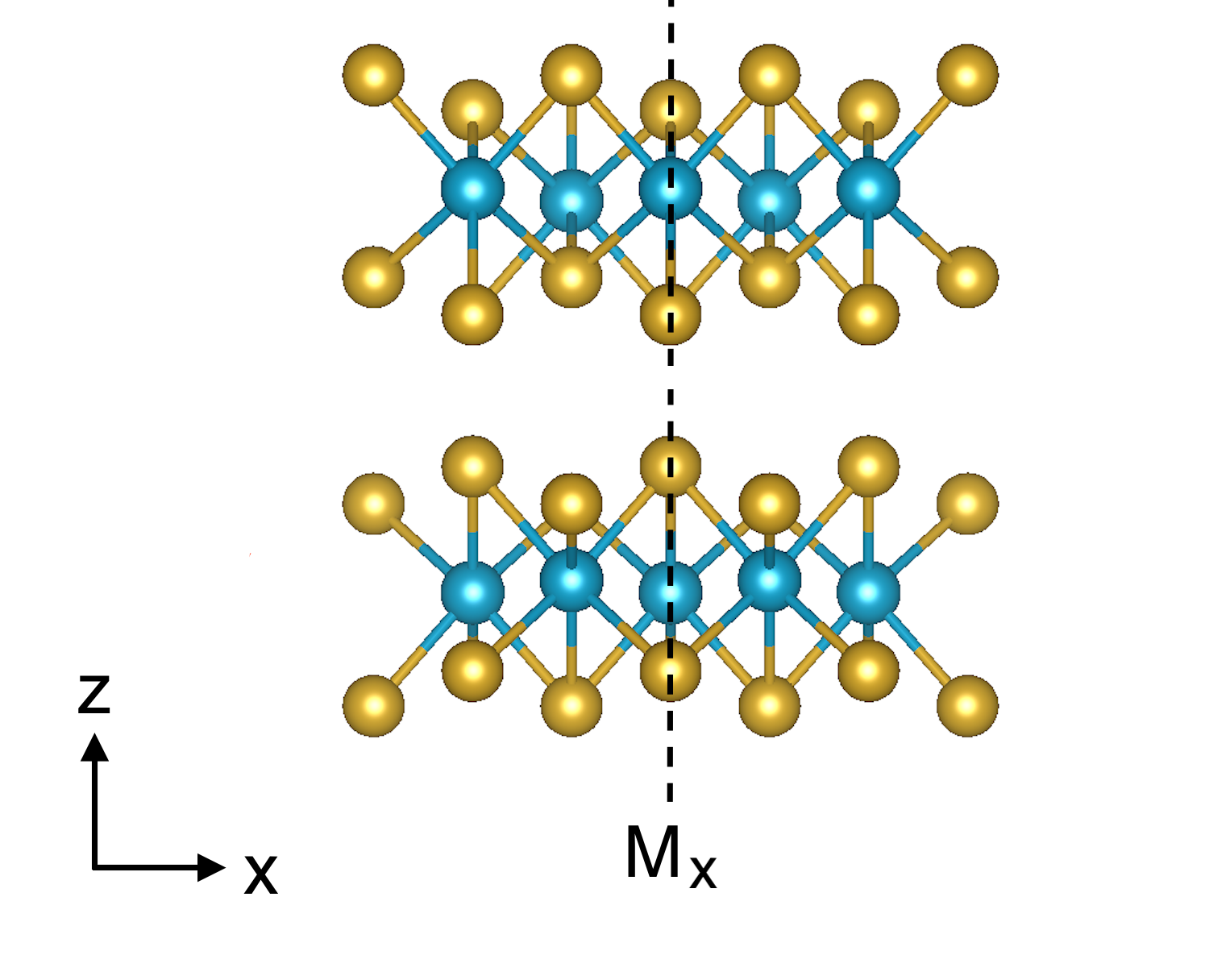}
    \end{minipage}
    \hfill
    \begin{minipage}{0.49\linewidth}
        \includegraphics[width=1\linewidth]{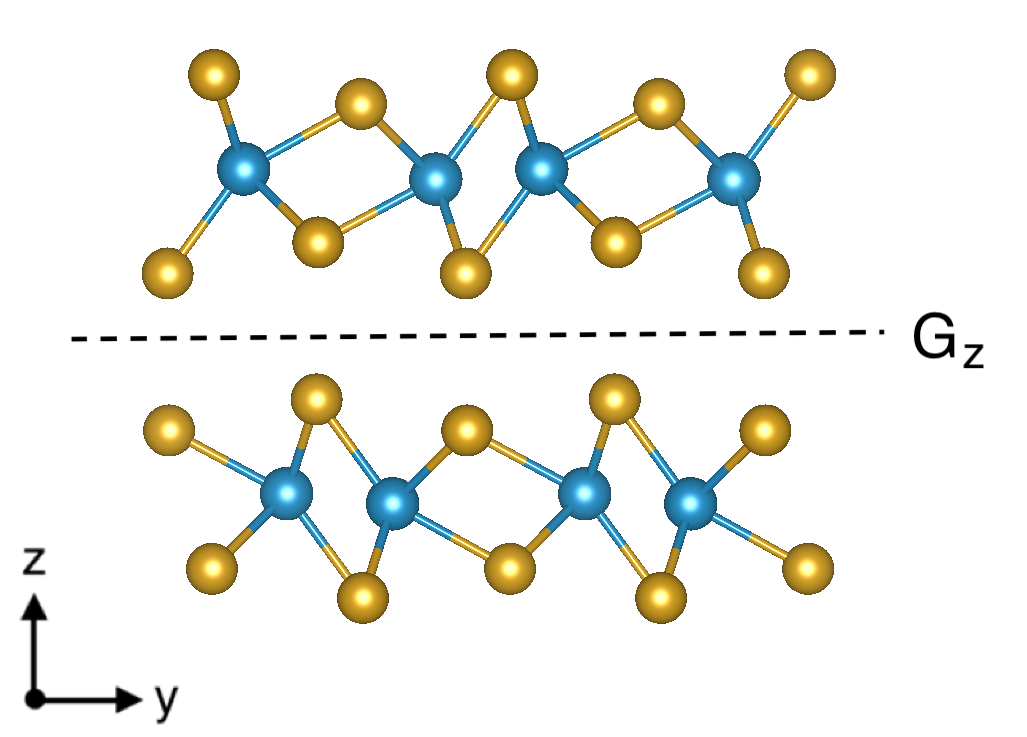}
    \end{minipage}
    \begin{picture}(0,0)
    \put(-110,90){\textbf{a)}}
    \put(0,90){\textbf{b)}}
    \end{picture}
    
    \caption{\textbf{Symmetries of the non-ferroelectric 1T$^\prime(^*)$ structure.} (a) Mirror plane $M_x$ (corresponding to the $yz$ plane).   (b) Glide-plane symmetry $G_z$, consisting of a reflection across the $xy$ plane combined with a fractional translation along the $y$ direction.}
    \label{fig:NFE_symm}
\end{figure}

\begin{figure}
    \centering
    
    \begin{minipage}{0.45\linewidth}
        \includegraphics[width=1.1\linewidth]{FE_XZ.png}
    \end{minipage}
    \hfill
    \begin{minipage}{0.53\linewidth}
        \includegraphics[width=1\linewidth]{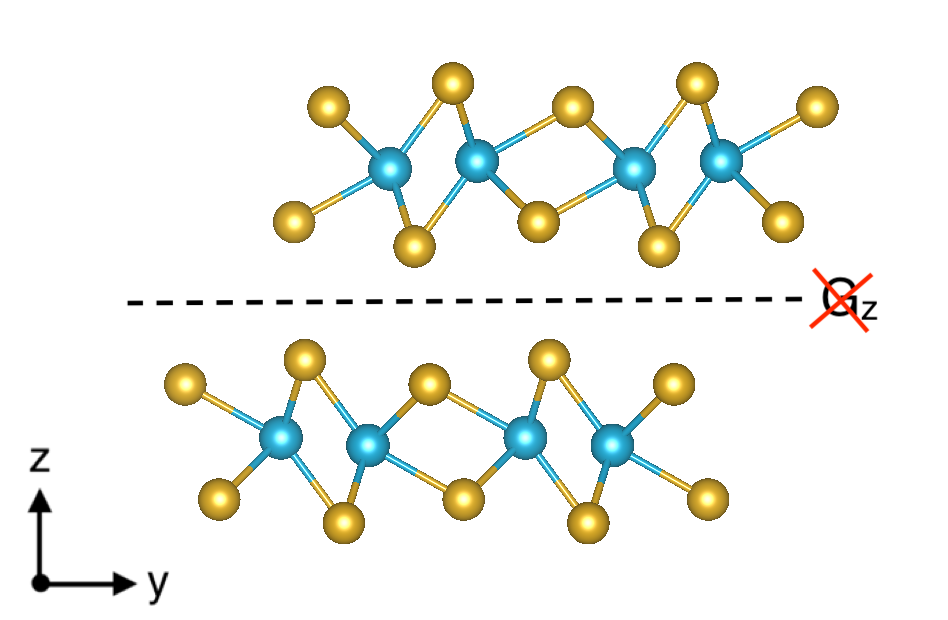}
    \end{minipage}
    \begin{picture}(0,0)
    \put(-110,90){\textbf{a)}}
    \put(0,90){\textbf{b)}}
    \end{picture}
    
    \caption{\textbf{Symmetries of the ferroelectric T$_\mathrm{d}$ structure.} (a) Mirror plane $M_x$. (b) Illustration of the broken glide-plane symmetry $G_z$. }
    \label{fig:FE_symm}
\end{figure}

The WTe$_2$ bilayer exhibits a $yz$ mirror plane, denoted as $M_x$, in both the NFE 1T$^\prime(^*)$ and FE T$_\mathrm{d}$ phases, as illustrated in \fref{fig:NFE_symm}a and \fref{fig:FE_symm}a, respectively. When HHG is driven by a linearly polarized field, whose polarization lies within this mirror plane (e.g., $y$ polarization), the mirror symmetry $M_x$ is preserved. As a consequence, harmonic emission perpendicular to the mirror plane--i.e., along the $x$ direction--is forbidden for all harmonic orders. This behaviour is clearly observed in \fref{fig:HHG_XY}c and f, where no harmonic response is detected in the $x$ component, consistent with the constraints imposed by $M_x$ symmetry.

When the WTe$_2$ bilayer is driven by an electric field polarized along the $x$ direction, i.e., perpendicular to the mirror plane $M_x$, a constraint arises for the even-order harmonics: the even-harmonic response must vanish perpendicular to the mirror plane $M_x$. The underlying explanation is that the $M_x$ mirror symmetry requires the orthogonal $x$ polarization of the induced electric field to have opposite sign and equal amplitude every half driving cycle (when the sign of the driving field is reversed). Such a behaviour is incompatible with even-order harmonics, whose dipole should be identical in both sign and amplitude every half driving cycle. Consequently, even harmonics polarized along $x$ are symmetry-forbidden, in agreement with the results shown in \fref{fig:HHG_XY}.

The key structural difference between the FE and NFE phases, relevant for the present analysis, is the presence of an $xy$ glide-plane symmetry, denoted by $G_z$, in the NFE 1T$^\prime(^*)$ phase (\fref{fig:NFE_symm}b), which is absent in the FE T$_\mathrm{d}$ phase. This glide operation involves a reflection across the $xy$ plane and a translation of half the lattice constant along the $y$ axis. As only the point-group component of the symmetry operation enters the HHG selection rules, $G_z$ is effectively equivalent to a mirror symmetry $M_z$ acting on the harmonic emission. Consequently, the presence of $G_z$ in the 1T$^\prime(^*)$ structure forbids the emission of $z$-polarized harmonics in the NFE phase.

In contrast, the breaking of $G_z$ in the FE T$_\mathrm{d}$ structure (\fref{fig:FE_symm}b) lifts this restriction, allowing for the generation of $z$-polarized harmonics. These symmetry-derived selection rules are consistent with the TDDFT results shown in \fref{fig:HHG_XY}a–f, where the $z$-polarized harmonics suppressed by approximately four orders of magnitude in the NFE phase compared to the FE phase. The residual, nonzero $z$-polarized response in the NFE structure can be attributed to the fact that the $G_z$ symmetry is not perfectly realized in the 1T$^\prime(^*)$ geometry obtained from the NEB calculations.

In \fref{fig:SYM_HHG}b, we quantify the deviation from the $G_z$ glide plane symmetry along the FE switching pathway through a mean absolute displacement measure (solid red line) and show that it directly correlates with the FE polarization along the $z$ axis, as well as the yield of $z$-polarized harmonics, depicted in \fref{fig:SYM_HHG}a.

From this analysis, we conclude that the suppression of $z$-polarized harmonics in the NFE phase is dictated by symmetry and can be directly attributed to the presence of the glide-plane symmetry $G_z$ in the 1T$^\prime(^*)$ structure. Conversely, in the FE T$_\mathrm{d}$ phase, the breaking of $G_z$ is intrinsically linked to the emergence of FE polarization. As a result, polarization-resolved HHG provides an indirect yet robust probe of the FE polarization in bilayer WTe$_2$.

For completeness, we demonstrate that the symmetry-derived HHG selection rules also apply to monolayer WTe$_2$. In contrast to the bilayer structures, the monolayer exhibits inversion symmetry $I$ and a two-fold rotational symmetry around the $x$ axis, $C_2(x)$. The inversion symmetry maps into the suppression of even harmonics in Figs. \ref{fig:HHG_XY}h,i. Moreover, we can distinguish two scenarios depending on the orientation of the driving polarization relative to the $x$ axis.  If the driving field is polarized along $x$, the system preserves its high symmetry, and all the harmonics are also polarized along $x$ (\fref{fig:HHG_XY}h). Contrarily, for a driving polarization along $y$ or $z$, the symmetry is reduced and superpositions of $y$ and $z$ components arise (\fref{fig:HHG_XY}i), consistent with the underlying symmetries of monolayer WTe$_2$.

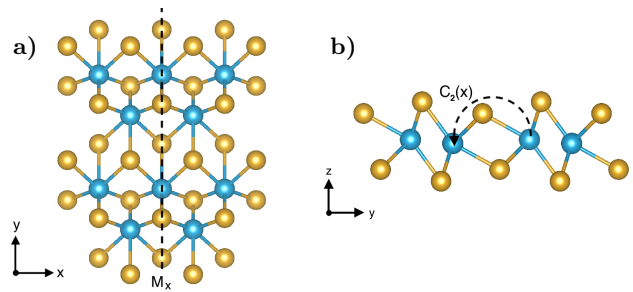
\begin{figure}
    \centering
    \begin{minipage}{0.45\linewidth}
        \includegraphics[width=1\linewidth]{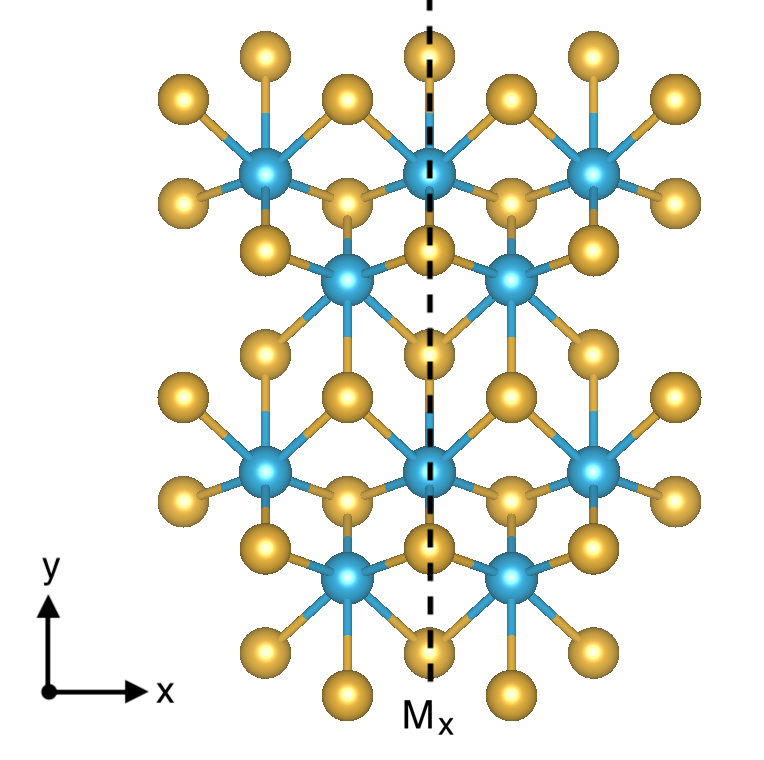}
    \end{minipage}
    \hfill\begin{minipage}{0.5\linewidth}
        \includegraphics[width=1\linewidth]{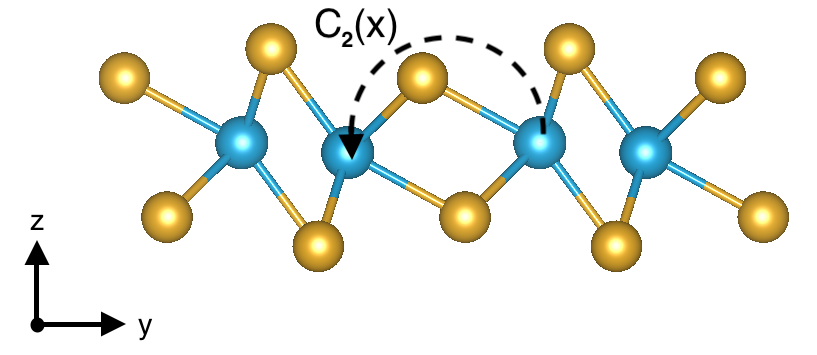}
    \end{minipage}
    \begin{picture}(0,0)
    \put(-240,40){\textbf{a)}}
    \put(-120,40){\textbf{b)}}
    \end{picture}
    \caption{\textbf{Symmetries of monolayer WTe$_2$.} (a) $xy$-plane of the monolayer highlighting the mirror symmetry $M_x$. (b) The monolayer additionally shows a two-fold rotational symmetry $C_2(x)$ around the $x$-axis.}
    \label{fig:monolayer_symm}
\end{figure}

\subsection*{Incorporating the lattice dynamics}
Having established the HHG response for static ionic configurations along the FE switching pathway, we now turn to the pump–probe scheme introduced in \fref{fig:experiment}a. In this setup, inspired by the experiments in Refs.~\onlinecite{sie2019ultrafastdyn,Zhang2019switch} and simulations in Refs.~\onlinecite{Yang2022_tddft,guan2021weyldyn}, a THz pump pulse excites coherent phonons in bilayer WTe$_2$, in particular an interlayer shear mode at \unit[0.24]{THz}, characterized by ionic motion along the $y$ direction of the FE switching pathway. A time-delayed near-infrared (NIR) probe pulse subsequently drives HHG, thereby probing the lattice structure and FE polarization, as discussed above.

\begin{figure}
    \centering
    \includegraphics[width=1\linewidth]{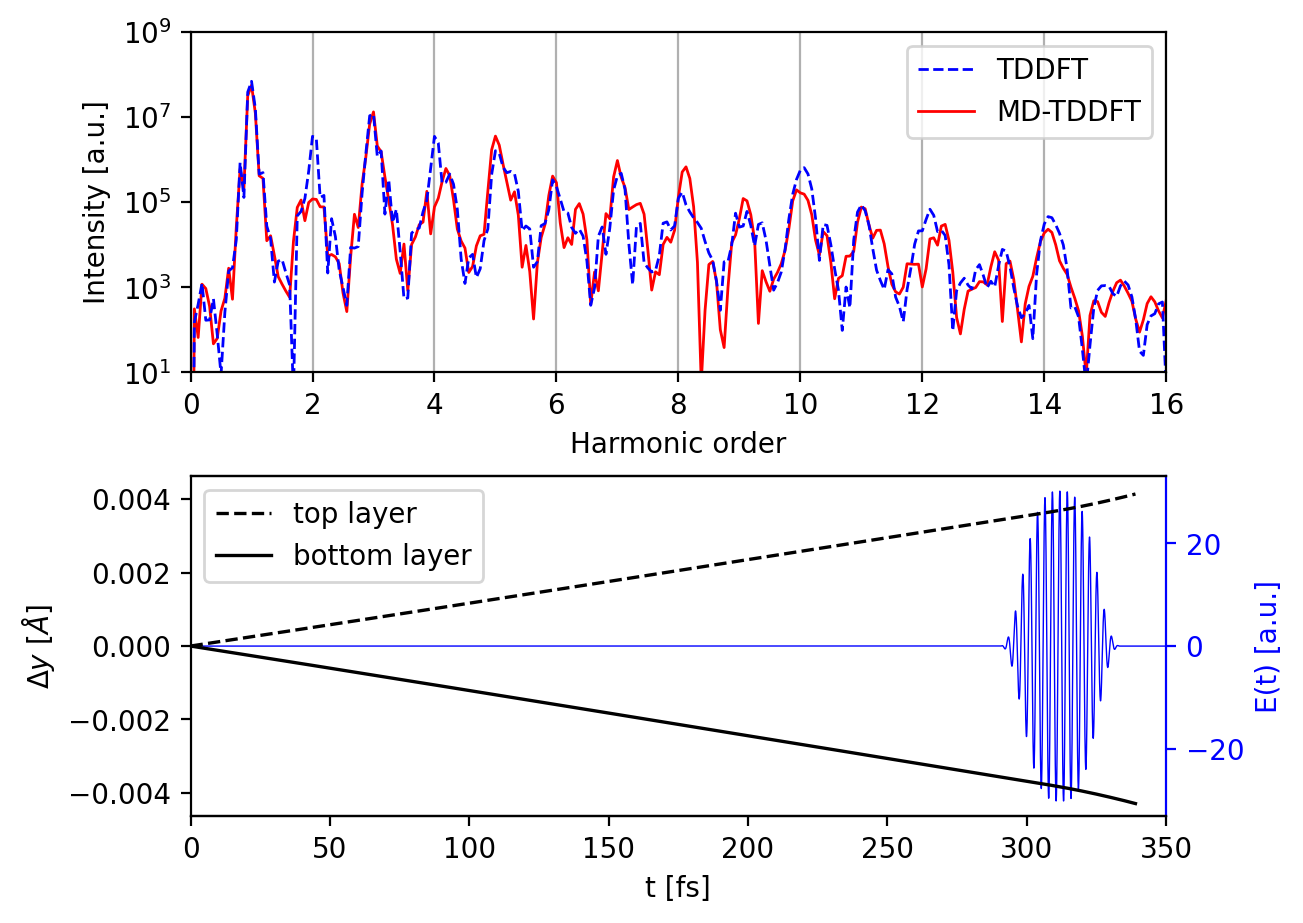}
    \begin{picture}(0,0)
    \put(-120,180){\textbf{a)}}
    \put(-120,100){\textbf{b)}}
    \end{picture}
    \caption{\textbf{Results of the MD-TDDFT simulations.} (a) Comparison of the total HHG spectrum obtained from coupled electron–ion dynamics (MD-TDDFT, red) with that from fixed ionic positions corresponding to the T$_\mathrm{d}$ phase of bilayer WTe$_2$ (TDDFT, blue). (b) Time evolution of the ionic displacements during the first \unit[350]{fs} following initialization of the interlayer shear mode. Shown are the displacements $\Delta y$ of the centers of mass of the top layer (dashed black line) and bottom layer (solid black line).  After \unit[300]{fs}, a 16-cycle \unit[800]{nm} probe pulse is applied, and the resulting HHG spectrum is shown in (a).}
    \label{fig:dynamic}
\end{figure}

To simulate the coupled electron–ion dynamics, we employ TDDFT combined with molecular dynamics (MD-TDDFT), as implemented in the Octopus code\cite{Octopus2020}. Within the Ehrenfest approximation\cite{alonso2008,Ehrenfest}, the electronic degrees of freedom are treated fully quantum mechanically within TDDFT, while the ions evolve along classical trajectories. The forces acting on the ions are obtained from the time-dependent electronic density in the presence of the external laser field.
The phononic excitation is modeled by directly initializing the interlayer shear mode starting from the ground-state T$_\mathrm{d}$ structure of bilayer WTe$_2$. This is achieved by assigning appropriate initial ionic velocities, rather than explicitly simulating the interaction with the THz pump pulse, thereby substantially reducing the computational cost. Further technical details of the MD-TDDFT simulations are provided in the Methods section.

Once the dynamics is initiated, the ions evolve according to the interlayer shear mode, and the coupled electron–ion system is propagated in time. The time evolution of the ionic displacements is depicted in \fref{fig:dynamic}b. After an initial propagation period, an \unit[800]{nm} probe pulse is applied, analogously to the simulations with frozen ionic configurations discussed above. Owing to the high computational cost, only a fraction of the full cycle of the \unit[0.24]{THz} shear mode (with a period of approximately \unit[4]{ps}) is simulated. In \fref{fig:dynamic}b, an $x$-polarized 16-cycle NIR probe pulse is applied after \unit[300]{fs}. The resulting HHG spectrum is shown in \fref{fig:dynamic}a (red) and compared with the spectrum obtained for the FE T$_\mathrm{d}$ phase with frozen ions (blue). We note that \fref{fig:dynamic}a displays the total HHG spectrum and is not polarization-resolved. A direct comparison reveals only minor differences, indicating that the spectra with and without ionic motion are nearly identical.

This observation suggests that ionic motion and electron–phonon coupling have a negligible impact on the HHG response in bilayer WTe$_2$ under the present conditions. This behaviour can be understood from the separation of time scales: the interlayer shear mode at \unit[0.24]{THz} evolves on picosecond time scales (with a period of approximately \unit[4]{ps}), whereas the electronic dynamics underlying HHG occurs on femtosecond time scales. As a result, the electronic response effectively follows the instantaneous ionic configuration, and the HHG process can be accurately described within a frozen-ion approximation. Consequently, the HHG results obtained for static ionic configurations, discussed above, remain applicable in the presence of lattice dynamics.
Finally, we note that accessing the NFE intermediate state requires excitation with intense THz pulses (7.5~MV/cm in Ref.\onlinecite{sie2019ultrafastdyn}), which drive anharmonic phonon dynamics that are not captured within the present simulations.

\section{Discussion}
We have investigated HHG in bilayer WTe$_2$, a prototypical 2D sliding ferroelectric with out-of-plane polarization. Our results demonstrate that HHG can serve as an all-optical probe of FE polarization. 
In particular, $z$-polarized harmonics--aligned with the out-of-plane FE polarization--are emitted only in the presence of FE order when the material is driven by an in-plane ($x$ or $y$) field.

We showed that this behaviour originates from symmetry constraints: in the NFE 1T$^\prime(^*)$ phase, a glide-plane symmetry $G_z$ forbids $z$-polarized emission, whereas its breaking in the FE T$_\mathrm{d}$ phase lifts this restriction. Since the emergence of FE polarization in bilayer WTe$_2$ is intrinsically linked to the breaking of $G_z$, polarization-resolved HHG provides a direct, symmetry-based probe of this order parameter. Moreover, the yield of the $z$-polarized harmonics scales with the degree of glide-plane symmetry breaking, enabling a quantitative connection between the HHG signal and the FE polarization along the switching pathway.

We further demonstrated that the Jones matrix formalism\cite{zhang2024jonesmatrix} consistently captures the symmetry-dependent selection rules governing HHG in both the FE T$_\mathrm{d}$ and NFE 1T$^\prime(^*)$ phases, as well as in the WTe$_2$ monolayer. This highlights the general applicability of the approach as a unifying framework for understanding symmetry-resolved HHG in solids.

More broadly, our results indicate that the connection between FE polarization and broken lattice symmetries provides a general principle for detecting ferroelectric order with HHG. In particular, a finite polarization requires the absence of symmetry operations, such as mirror or glide planes, that reverse the polar axis. This symmetry breaking opens otherwise forbidden harmonic emission channels, allowing polarization-resolved HHG to act as a direct optical fingerprint of the FE state.   We therefore expect this mechanism to be applicable beyond WTe$_2$, extending to a broad class of ferroelectric and polar materials. Our work establishes HHG as a sensitive, all-optical probe of ferroelectricity and lattice-symmetry breaking, and highlights its potential for tracking ultrafast structural dynamics, interlayer sliding, and polarization switching in quantum materials.

\section{Methods}
{\bf DFT.}
The DFT calculations, including structural relaxations and nudged elastic band (NEB) calculations, were performed using the Vienna ab-initio simulation package (VASP)\cite{PhysRevBVASP}. The exchange–correlation functional was treated within the Perdew–Burke–Ernzerhof functional (PBE\cite{PhysRevLettPBE}) generalized gradient approximation, neglecting spin polarization and spin–orbit coupling. Core states were described using projector augmented-wave pseudopotentials\cite{Kresse1999}. 
A $8 \times 8 \times 1$ Monkhorst–Pack $k$-point mesh and a plane-wave cutoff energy of \unit[500]{eV} were used, with an energy convergence criterion of \unit[10$^{-8}$]{eV}. Van der Waals interactions were included via the DFT-D3 method with Becke–Johnson damping\cite{dft-d3} for all structural relaxations and NEB calculations.

Bilayer T$_\mathrm{d}$-WTe$_2$ adopts a non-centrosymmetric orthorhombic structure (space group $Pnm2_1$), enabling out-of-plane ferroelectric polarization. The optimized lattice parameters are $a=\unit[3.461]{\AA}$, $b = \unit[6.287]{\AA}$, and $c = \unit[10.106]{\AA}$. A vacuum spacing of \unit[17.6]{\AA} was applied along the $z$ direction. Atomic coordinates are provided in the Supplementary Information.

For the final NEB state, the top layer was displaced by \unit[0.72]{\AA} along $y$ and relaxed to obtain the  structure with reversed polarization \cite{yang2018dft}. Structural relaxations employed a conjugate gradient algorithm with an energy convergence threshold of \unit[0.005]{eV}. NEB calculations were performed at fixed cell shape and volume with a convergence criterion of \unit[0.01]{eV}. The nine intermediate images were subsequently recalculated including dipole corrections along $z$ to extract the interlayer dipole moment.\\

{\bf TDDFT.}
Time-dependent density functional theory (TDDFT) calculations were performed using Octopus\cite{Octopus2020}, which describes electron dynamics on a real-space grid by solving the time-dependent Kohn–Sham (KS) equations:
\begin{equation}
i\partial_t \ket{\Psi_{nk}(r,t)} = \Bigl[ \frac{1}{2} \left(-i\nabla - \frac{1}{c} \kvec{A}(t)\right)^2 + v_s(r,t) \Bigr]\ket{\Psi_{nk}(r,t)}.
\label{eq:TDKS}
\end{equation}
Here, $\ket{\Psi_{nk}(r,t)}$ are the KS single-particle wave functions for band $n$ and k-point $k$.  $\kvec{A}(t)$ denotes the vector potential in the velocity gauge, $c$ the speed of light, and $v_s(r,t)$ the time-dependent KS potential, which includes the ionic Coulomb potential, the Hartree term, and the exchange–correlation potential treated within the adiabatic local density approximation (ALDA).
Starting from the ground state electronic occupations, time propagation was carried out using the enforced time-reversal symmetry (etrs) scheme \cite{octopus:etrs}, in which the time-evolution operator is approximated on a discretized time grid. A time step of $\Delta t = \unit[0.05]{a.u.} =\unit[2.065]{fs}$ was employed throughout the simulations.

The vector potential $\kvec{A}(t)$ is related to the electric field via $-\partial_t \kvec{A}(t) = c\kvec{E}(t)$ and is defined as 
\begin{equation}
    \kvec{A}(t) = -f(t) \frac{E_0}{\omega}c\ \sin(\omega t)\hat{\kvec{e}},
    \label{eq:laser}
\end{equation}
where $E_0$ denotes the electric field amplitude, $\omega$ the driving frequency, and $\hat{\kvec{e}}$ a unit vector specifying the polarization direction.  

The temporal envelope $f(t)$ was taken as a normalized function of the form\cite{Neufeld2019}
\begin{equation}
f(t) = \sin\left(\frac{\pi t}{T_p}\right)^{\left(\frac{\left|\pi\left(\frac{t}{T_p}-\frac{1}{2}\right)\right|}{\sigma}\right)},
\label{eq:envelope}
\end{equation}
where $\sigma = 0.75$, and $T_p$ denotes the total pulse duration. In this work, $T_p$ was chosen as $T_p = 16T$, with $T = 2\pi/\omega$ corresponding to a single cycle of the carrier frequency $\omega$.
In the simulations, we used a wavelength $\lambda = 2\pi c/\omega = \unit[800]{nm}$ and a peak intensity of $E_0 = \unit[1]{TW/cm^2}$.
The TDDFT calculations were performed for the relaxed structures obtained from VASP, including intermediate configurations along the NEB pathway. The exchange-correlation functional was treated within LDA. Spin degrees of freedom and spin-orbit coupling were neglected. Core electrons were described within the frozen-core approximation using norm-conserving pseudopotentials.\cite{hartwigsen1998} A $7\times5\times1$ k-grid and a real-space grid spacing of \unit[0.3]{\AA} were employed throughout. 

A direct output of the TDDFT simulations is the time-dependent current density
\begin{equation}
\label{eq:curr}
\mathbf{j}(\mathbf{r},t) = \frac{1}{2} \sum_{nk} \left[ \Psi_{nk}^*(\mathbf{r},t) \left( -i\nabla - \frac{\mathbf{A}(t)}{c}  \right) \Psi_{nk}(\mathbf{r},t) + c.c. \right],
\end{equation}
from which the total current expectation value $\mathbf{J}(t)$ is obtained by integration over the unit cell of volume $V$
\begin{equation}
\mathbf{J}(t) = \frac{1}{V} \int_V \mathbf{j}(\mathbf{r},t)d^3r.
\end{equation}
The polarization-resolved HHG spectrum was calculated as the Fourier transform of the derivative of the current components $J_i(t)$ along each direction $i \in (x,y,z)$
\begin{equation}
    \text{I}_i(\omega) = \left|\int dt\partial_t J_i(t)e^{-i\omega t}\right|^2,
    \label{eq:HHG}
\end{equation}
which was evaluated numerically using a fast Fourier transform.

The harmonic yield shown in \fref{fig:SYM_HHG}a is obtained by applying a moving average to the HHG spectra along the frequency axis for each intermediate structure. This smoothing enhances the visibility of the harmonic trends without affecting their spectral positions. The width of the moving average is chosen as 0.237 times the fundamental frequency, corresponding to a fraction of the harmonic spacing. The resulting spectra are subsequently interpolated across the intermediate structures along the transition pathway to obtain a smooth representation.\\

{\bf Symmetry analysis.}

To gain insight into symmetry-dependent HHG selection rules, we employ the Jones matrix formalism introduced in Ref.~\onlinecite{zhang2024jonesmatrix}. This approach enables the determination of dipole-allowed harmonic orders, $n_H$, for a given polarization. The Jones matrix $J(n_H)$ for HHG from crystalline solids is given by\cite{zhang2024jonesmatrix} 
\begin{equation}
    J(n_H) = \sum_{l'=1}^{N'}(R') \sum_{l=1}^{N} \exp(-i2\pi n_H\frac{l}{N})(R)^l,
\label{eq:JonesMatrix} 
\end{equation}
where $n_H$ is the harmonic order, and $R$ and $R'$ represent the shared orthogonal  transformations of the external field and the crystal structure.
The integers $N$ and $N'$ correspond to the orders of the symmetry operations $R$ and $R'$, respectively, such that $R^N = E$ and $(R')^{N'} = E$, with $E$ being the identity operator. For instance, mirror symmetry or two-fold rotational symmetry corresponds to $N = 2$.
The symmetry operation $R'$ leaves both the crystal structure and the external field invariant. In contrast, $R$ preserves the crystal symmetry while inducing a temporal translation of the driving field, $R\mathbf{A}(t) = \mathbf{A}(t + T_0/N)$, where $T_0$ is the fundamental period of the field. In the following, we explicitly construct the Jones matrices for the crystal symmetries and driving-field polarizations considered in this work.

\begin{figure}
    \centering
    \begin{minipage}{0.53\linewidth}
    \vspace{0.5cm}
        \includegraphics[width=1\linewidth]{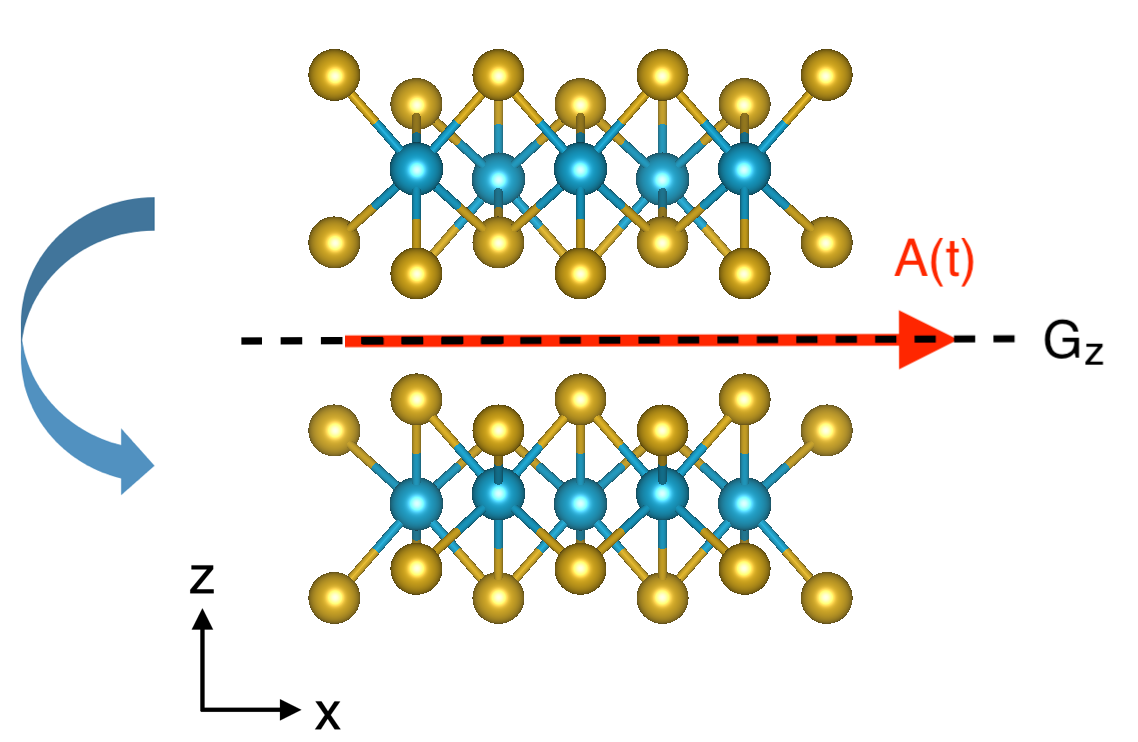}
    \end{minipage}
    \hfill
    \begin{minipage}{0.43\linewidth}
        \includegraphics[width=1\linewidth]{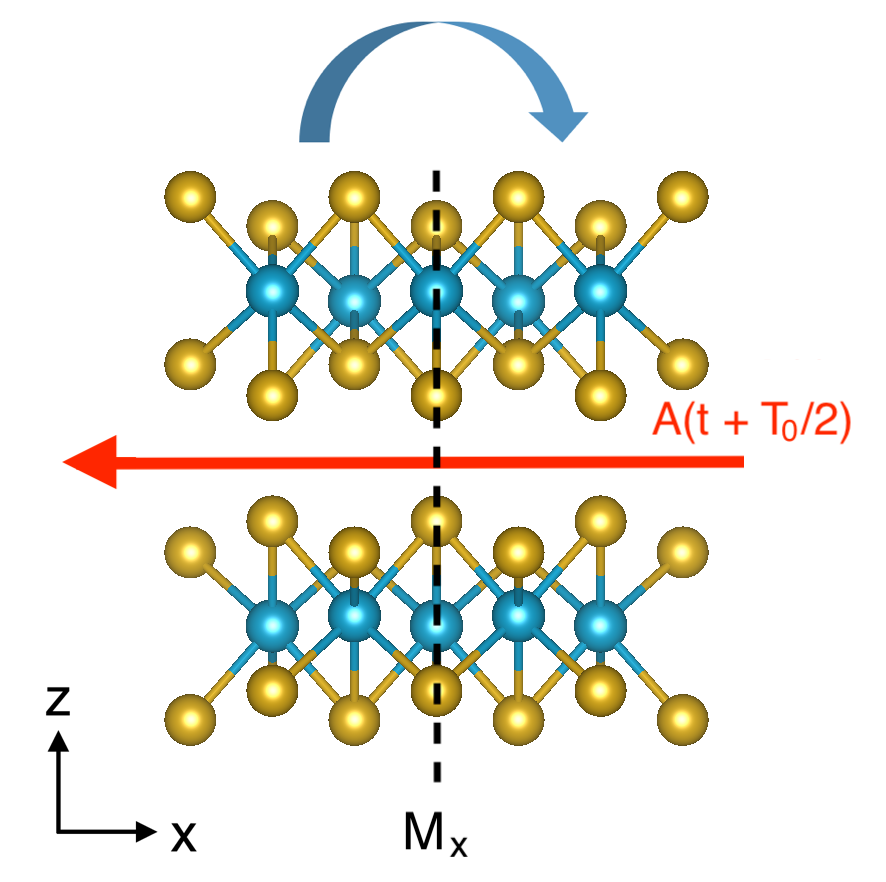}
    \end{minipage}
    \begin{picture}(0,0)
    \put(-240,40){\textbf{a)}}
    \put(-120,40){\textbf{b)}}
    \put(-180,40){$R' = G_z$}
    \put(-30,40){$R = M_x$}
    \end{picture}
    \caption{
    \textbf{Shared symmetries of the NFE 1T$^\prime(^*)$ structure and an $x$-polarized external field.} (a) The glide-plane symmetry $G_z$ leaves both the crystal structure and the external field invariant. (b) The mirror symmetry $M_x$ maps the external field onto itself with a temporal shift of $T_0/2$, corresponding to a reversal of its direction.}
    \label{fig:symmetry_analysis}
\end{figure}
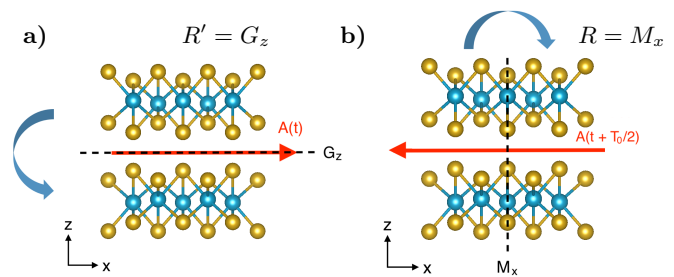

We begin with the NFE 1T$^\prime(^*)$ structure of bilayer WTe$_2$. As illustrated in Fig.~\ref{fig:symmetry_analysis}, this structure possesses a mirror plane perpendicular to the $x$ axis (the $yz$ plane), denoted as $M_x$, and a glide plane parallel to the $xy$ plane, denoted as $G_z$. The glide-plane symmetry comprises a reflection across the $xy$ plane, corresponding to $z \rightarrow -z$, combined with a fractional translation of \unit[3.14]{\AA} along the $y$ direction. Since only the orthogonal (point-group) component of the symmetry operation enters the Jones matrix formalism, the translational part of the glide operation is not considered.

For a driving field linearly polarized along $x$, the only symmetry operation that leaves both the crystal structure and the external field invariant is the point-group component of $G_z$. Accordingly, as shown in \fref{fig:symmetry_analysis}a, we obtain
\begin{equation}
R' = G_z = \Bigl(\begin{smallmatrix}
1 &  &  \\
 & 1 &  \\
 &  & -1 \\
\end{smallmatrix}\Bigr). 
\label{eq:R'}
\end{equation}
In contrast, the symmetry operation that preserves the crystal structure while shifting the external field by $T_0/2$ is the mirror operation $M_x$. As illustrated in \fref{fig:symmetry_analysis}b, this yields
\begin{equation}
R = M_x = \Bigl(\begin{smallmatrix}
-1 &  &  \\
 & 1 &  \\
 &  & 1 \\
\end{smallmatrix}\Bigr).
\label{eq:R}
\end{equation}
Substituting Eqs.~\eqref{eq:R} and \eqref{eq:R'} into the Jones matrix expression given in Eq.~\eqref{eq:JonesMatrix}, we obtain
\begin{equation}
    J(n_H)=\Biggl(\begin{smallmatrix} 1- e^{-i\pi n_H}&&\\&1 + e^{-i\pi n_H} &\\&&0\\\end{smallmatrix}\Biggr).
\label{eq:Jones_x}
\end{equation}
We therefore conclude that, for the NFE 1T$^\prime(^*)$ structure driven by an $x$-polarized field, only $x$-polarized odd harmonics ($n_H = 2n+1$) and $y$-polarized even harmonics ($n_H = 2n$) are allowed, while harmonics polarized along $z$ are symmetry-forbidden. These conclusions are  consistent with the harmonics observed in \fref{fig:HHG_XY}e.

Next, we analyze the symmetries shared by the NFE 1T$^\prime(^*)$ structure and a $y$-polarized external field. In this case, the laser polarization lies within the $yz$ mirror plane of the lattice; consequently, both $M_x$ and $G_z$ leave the crystal structure and the external field invariant. In contrast, there is no symmetry operation of the lattice that translates the external field by $T_0/2$. We therefore obtain
\begin{equation}
R' = G_zM_x = \Bigl(\begin{smallmatrix}
-1 &  &  \\
 & 1 &  \\
 &  & -1 \\
\end{smallmatrix}\Bigr) \;\; \mathrm{and} \;\;
R = E = \Bigl(\begin{smallmatrix}
1 &  &  \\
 & 1 &  \\
 &  & 1 \\
\end{smallmatrix}\Bigr).
\label{eq:NFE_y_R'}
\end{equation}
Substituting $R$ and $R'$ into Eq.~\eqref{eq:JonesMatrix}, we obtain the Jones matrix
\begin{equation}
    J(n_H)\propto\Bigl(\begin{smallmatrix} 0&&\\&1&\\&&0\\\end{smallmatrix}\Bigr).
\label{eq:Jones_NFE_y}
\end{equation}
This result implies that $y$-polarized harmonics of even and odd order are allowed, whereas $x$- and $z$-polarized harmonics are symmetry-forbidden. These predictions are consistent with the harmonics observed in \fref{fig:FE_symm}f.\\

The FE T$_\mathrm{d}$ structure of bilayer WTe$_2$ retains the $yz$ mirror plane, denoted by $M_x$, while the glide-plane symmetry $G_z$ is broken, as shown in Fig.~\ref{fig:FE_symm}. Consequently, for an $x$-polarized driving field, there exists no orthogonal transformation that leaves both the crystal structure and the external field invariant. Instead, the mirror symmetry $M_x$ maps the external field as $\mathbf{A}(t) \rightarrow \mathbf{A}(t + T_0/2)$. The relevant symmetry operations are therefore
\begin{equation}
R' = E = \Bigl(\begin{smallmatrix}
1 &  &  \\
 & 1 &  \\
 &  & 1 \\
\end{smallmatrix}\Bigr)  \;\; \mathrm{and} \;\;
R = M_x = \Bigl(\begin{smallmatrix}
-1 &  &  \\
 & 1 &  \\
 &  & 1 \\
\end{smallmatrix}\Bigr),
\label{eq:NFE_y_R'}
\end{equation}
which yield the Jones matrix
\begin{equation}
    J(n_H)=\Biggl(\begin{smallmatrix} 1- e^{-i\pi n_H}&&\\&1 + e^{-i\pi n_H} &\\&&1 + e^{-i\pi n_H}\\\end{smallmatrix}\Biggr).
\label{eq:Jones_x_FE}
\end{equation}
This result indicates that, for an $x$-polarized driving field, odd harmonics are emitted with $x$ polarization, while even harmonics appear in the $y$ and $z$ components. These predictions are confirmed by the TDDFT results shown in Fig.~\ref{fig:SYM_HHG}b.

For a $y$-polarized external field, the mirror symmetry $M_x$ leaves both the crystal structure and the field invariant, while no lattice symmetry exists that induces a temporal shift of $T_0/2$. The corresponding symmetry operations are therefore
\begin{equation}
R' = M_x = \Bigl(\begin{smallmatrix}
-1 &  &  \\
 & 1 &  \\
 &  & 1 \\
\end{smallmatrix}\Bigr) \;\; \mathrm{and} \;\;
R = E = \Bigl(\begin{smallmatrix}
1 &  &  \\
 & 1 &  \\
 &  & 1 \\
\end{smallmatrix}\Bigr),
\label{eq:FE_y_R'}
\end{equation}
which lead to the Jones matrix 
\begin{equation}
J(n_H)\propto\Bigl(\begin{smallmatrix} 0&&\\&1&\\&&1\\\end{smallmatrix}\Bigr).
\label{eq:Jones_x_FE}
\end{equation}
Accordingly, both even- and odd-order harmonics are allowed with $y$ and $z$ polarization, while $x$-polarized harmonics are forbidden by symmetry. \\

The investigated monolayer structure exhibits, in addition to the mirror symmetry $M_x$, inversion symmetry $I$, and a two-fold rotational symmetry about the $x$ axis, denoted by $C_2(x)$. For an $x$-polarized driving field, both the crystal structure and the external field are invariant under the rotation $C_2(x)$. In contrast, inversion symmetry $I$ reverses the direction of the external field, corresponding to $\mathbf{A}(t) \rightarrow \mathbf{A}(t + T_0/2)$. The resulting symmetry operations are therefore  
\begin{equation}
R' = C_2(x) =  \Bigl(\begin{smallmatrix}
1 &  &  \\
 & -1 &  \\
 &  & -1 \\
\end{smallmatrix}\Bigr) \;\; \mathrm{and} \;\;
R = I = \Bigl(\begin{smallmatrix}
-1 &  &  \\
 & -1 &  \\
 &  & -1 \\
\end{smallmatrix}\Bigr),
\label{eq:mono_x_RR'}
\end{equation}
which lead to the Jones Matrix
\begin{equation}
    J(n_H)=\Biggl(\begin{smallmatrix} 1 - e^{-i\pi n_H}&&\\&0 &\\&&0 \\\end{smallmatrix}\Biggr)
\label{eq:Jones_x_mono}
\end{equation}
This result implies that only odd-order harmonics with $x$ polarization are allowed, in agreement with the TDDFT results shown in \fref{fig:HHG_XY}h.

For a $y$-polarized driving field, the symmetry operations shared by the monolayer and the external field are
 \begin{equation}
R' = M_x = \Bigl(\begin{smallmatrix}
-1 &  &  \\
 & 1 &  \\
 &  & 1 \\
\end{smallmatrix}\Bigr) \;\; \mathrm{and} \;\;
R = I = \Bigl(\begin{smallmatrix}
-1 &  &  \\
 & -1 &  \\
 &  & -1 \\
\end{smallmatrix}\Bigr),
\label{eq:mono_y_R'}
\end{equation}
which yield the Jones matrix
\begin{equation}
    J(n_H)=\Biggl(\begin{smallmatrix} 0&&\\&1 - e^{-i\pi n_H} &\\&&1 - e^{-i\pi n_H} \\\end{smallmatrix}\Biggr),
\label{eq:Jones_x_mono}
\end{equation}
Accordingly, only odd-order harmonics polarized along $y$ and $z$ are allowed, consistent with the TDDFT results.

Finally, we summarize all shared symmetries of the various crystal structures and laser polarizations, including driving fields polarized along $z$, in Table~\ref{tab:allowed_harmonics}.\\

\begin{table}[]
    \centering
    \caption{Shared symmetries $R$ and $R'$ for the investigated crystal structures and external field polarizations. The relevant symmetry operations include mirror symmetry $M_x$, glide-plane symmetry $G_z$, two-fold rotation $C_2(x)$, and inversion symmetry $I$.}
    \label{tab:shared_symmetries}
    \begin{tabular}{l|c|c|c|c|c|c}
        \hline
        Fundamental & \multicolumn{2}{c}{$x$} & \multicolumn{2}{c}{$y$} & \multicolumn{2}{c}{$z$}\\
        \hline
        Symmetry & \;\;\;$R$\;\;\; &\;\;\;$R'$\;\;\; & \;\;\;\;$R$\;\;\;\; & $R'$ & \;\;\;\;$R$\;\; \;&\; \;\;$R'$\;\;\; \\
        \hline
        T$_\mathrm{d}$ & $M_x$ & $E$ & $E$ & $M_x$ & $E$&$M_x$\\
        \hline
        1T$^\prime(^*)$ &$M_x$&$G_z$ &$E$&$M_x$ $G_z$ &$G_z$ & $M_x$\\ 
        \hline
        monolayer & $I$ & $C_2(x)$&$C_2(x)$ &$I$&$C_2(x)$ & $I$\\
        \hline
    \end{tabular}
\end{table}

The deviation from the glide-plane symmetry $G_z$, shown in Fig.~\ref{fig:SYM_HHG}b, is quantified using a mean absolute displacement measure. For each intermediate structure along the NEB pathway, the atomic coordinates are first reflected across the $xy$ plane (i.e., along the $z$ direction). The mirrored structure is then rigidly displaced so as to minimize its deviation from the original configuration. The residual distances are evaluated according to
\begin{equation}
D = \frac{1}{N} \sum_{i=1}^{N} \lvert \mathbf{r}_i - \mathbf{r}'_i \rvert,
\label{eq:RMSD}
\end{equation}
where $N$ is the number of atoms in the unit cell, $\mathbf{r}_i$ denotes the position of the $i$-th atom in the original structure, and $\mathbf{r}'_i$ the position of the corresponding atom in the mirrored and displaced configuration. For a perfectly glide-plane symmetric structure, $D$ vanishes. In the NFE 1T$^\prime(^*)$ phase, we obtain a residual value of \unit[0.0028]{\AA}.\\

{\bf MD-TDDFT.}
The molecular dynamics time-dependent density functional theory (MD-TDDFT) simulations were performed using Octopus\cite{Octopus2020}. The coupled electron-ion dynamics was treated within the Ehrenfest approximation\cite{Ehrenfest}, in which the ions evolve along classical trajectories while the electron dynamics is described quantum mechanically within TDDFT. The forces acting on the ions are computed at each time step from the time-dependent electronic density.

The T$_\mathrm{d}$ ground-state structure of bilayer WTe$_2$ was used as the initial configuration for the time propagation. To eliminate residual forces, the structure was relaxed until the atomic forces were below \unit[10$^{-6}$]{a.u}. The interlayer shear phonon mode was then initialized by assigning initial ionic velocities of \unit[6.5$\times10^{-6}$]{\AA/fs} along the corresponding eigenmode obtained from density functional perturbation theory (DFPT).

All calculations were performed using a $7\times5\times1$ $k$-point grid, LDA pseudopotentials, and a time step of \unit[0.05]{a.u}. The parameters of the time-delayed NIR pulse used to drive HHG were identical to those employed for the static ionic configurations described above.

\section{Data availability}

\clearpage

%

\section{Acknowledgments} 
We thank Nicolas Tancogne-Dejean, Kazuhiro Yabana, Dongbin Shin, and Ofer Neufeld for helpful discussions.
This work was financially supported
by the Austrian Science Fund (FWF)
grant 10.55776/V988, the European Research Council (ERC-2024-SyG-101167294; UnMySt), the Cluster of Excellence Advanced Imaging of Matter (AIM), Grupos Consolidados y Alto Rendimiento UPV/EHU, Gobierno Vasco (IT1453-22).
We acknowledge support from the Max Planck-New York City Center for
Non-Equilibrium Quantum Phenomena. The Flatiron Institute is a division of the Simons Foundation.  A.H. acknowledges the support from the postdoctoral fellowship programme of the Alexander von Humboldt Foundation.
The computational results have been achieved using the Austrian Scientific Computing (ASC) infrastructure.

\section{Author contributions}
A.G. conceived, designed, and led the project. E.G. performed the calculations and symmetry analysis under A.G.’s supervision. A.H. contributed to data analysis and scientific discussions. A.G., E.G., and A.H. drafted the manuscript. All authors discussed the results and revised the manuscript.

\section{Competing interests}
The authors declare no competing interests.

\end{document}